\documentclass[aps,prc,twocolumn,amsmath,amssymb,superscriptaddress,nofootinbib,showpacs,showkeys]{revtex4-1}
\usepackage{dcolumn}
\usepackage{graphicx}
\usepackage{color}
\usepackage{multirow}
\usepackage{textcomp}
\usepackage{physics}
\usepackage{dsfont}
\usepackage{epstopdf}

\usepackage{relsize}
\begin{document}
	\newcommand {\nc} {\newcommand}
	\nc {\beq} {\begin{eqnarray}}
	\nc {\eeq} {\nonumber \end{eqnarray}}
	\nc {\eeqn}[1] {\label {#1} \end{eqnarray}}
\nc {\eol} {\nonumber \\}
\nc {\eoln}[1] {\label {#1} \\}
\nc {\ve} [1] {\mbox{\boldmath $#1$}}
\nc {\ves} [1] {\mbox{\boldmath ${\scriptstyle #1}$}}
\nc {\mrm} [1] {\mathrm{#1}}
\nc {\half} {\mbox{$\frac{1}{2}$}}
\nc {\thal} {\mbox{$\frac{3}{2}$}}
\nc {\fial} {\mbox{$\frac{5}{2}$}}
\nc {\la} {\mbox{$\langle$}}
\nc {\ra} {\mbox{$\rangle$}}
\nc {\etal} {\emph{et al.}}
\nc {\eq} [1] {(\ref{#1})}
\nc {\Eq} [1] {Eq.~(\ref{#1})}
\nc {\Refc} [2] {Refs.~\cite[#1]{#2}}
\nc {\Sec} [1] {Sec.~\ref{#1}}
\nc {\chap} [1] {Chapter~\ref{#1}}
\nc {\anx} [1] {Appendix~\ref{#1}}
\nc {\tbl} [1] {Table~\ref{#1}}
\nc {\Fig} [1] {Fig.~\ref{#1}}
\nc {\ex} [1] {$^{#1}$}
\nc {\Sch} {Schr\"odinger }
\nc {\flim} [2] {\mathop{\longrightarrow}\limits_{{#1}\rightarrow{#2}}}
\nc {\textdegr}{$^{\circ}$}
\nc {\inred} [1]{\textcolor{red}{#1}}
\nc {\inblue} [1]{\textcolor{blue}{#1}}
\nc {\IR} [1]{\textcolor{red}{#1}}
\nc {\IB} [1]{\textcolor{blue}{#1}}
\nc{\pderiv}[2]{\cfrac{\partial #1}{\partial #2}}
\nc{\deriv}[2]{\cfrac{d#1}{d#2}}
\title{Detailed Study of the Eikonal Reaction Theory for the breakup of one-neutron halo nuclei}
\author{C.~Hebborn}
\email{hebborn@frib.msu.edu}
\affiliation{Facility for Rare Isotope Beams, Michigan State University, East Lansing, MI 48824, USA}
\affiliation{Lawrence Livermore National Laboratory, P.O. Box 808, L-414, Livermore, California 94551, USA}
\affiliation{Physique Nucl\' eaire et Physique Quantique (CP 229), Universit\'e libre de Bruxelles (ULB), B-1050 Brussels}
\author{P.~Capel}
\email{pcapel@uni-mainz.de}
\affiliation{Institut f\"ur Kernphysik, Johannes Gutenberg-Universit\"at Mainz, D-55099 Mainz}
\affiliation{Physique Nucl\' eaire et Physique Quantique (CP 229), Universit\'e libre de Bruxelles (ULB), B-1050 Brussels}
\date{\today}
\begin{abstract}
\begin{description}
\item[Background] One-neutron removal reactions are used to study the single-particle structure of unstable nuclei, and in particular the exotic {halo nuclei}.
The Eikonal Reaction Theory (ERT) has been developed by Yahiro, Ogata and 
Minomo in Prog.\ Theor.\ Phys.\ {\bf 126}, 167 (2011) to include dynamical effects, which are missing in the usual eikonal description of these reactions.
Encouraging results have been obtained for total breakup cross sections in comparison to more elaborate reaction models.
\item[Purpose] We extend these comparisons to more {differential} breakup cross 
sections expressed as functions of the relative energy or parallel momentum {between  the core and  halo neutron}.
\item[Method] ERT predictions of these cross sections are compared to state-of-the-art calculations.
\item[Results] The hypotheses upon which the ERT is based are confirmed and their range of validity is made clearer.
The actual ordering of the evolution operators {affects ERT differential} cross sections and a specific choice leads to excellent agreement with the reference calculation.
Dynamical effects in the treatment of the neutron-target interaction can be significant in the parallel-momentum observable.
\item[Conclusions] The role of the different interactions in the dynamics 
of breakup reactions of one-neutron halo nuclei are better understood and 
improvements to the ERT are suggested.
\end{description}

\end{abstract}
\maketitle
%


\section{Introduction}\label{Introduction}

Due to their peculiar structure, halo nuclei have been studied intensively since their discovery in the mid-1980s~\cite{Tetal85a,T96}.  These nuclei exhibit a very large matter radius compared to their isobars.
This surprisingly large size is qualitatively understood by the decoupling of one or two valence nucleons from the nucleus thanks to their small separation energy. 
{They tunnel deep into the classically-forbidden region to form a diffuse} halo around a compact core that has the same properties---size and density---as a stable nucleus \cite{HJ87}. These nuclei are therefore often described as few-body objects: an inert core to which one or two nucleons are loosely bound.
Archetypical halo nuclei are $^{11}\rm Be$, seen as $^{10}\rm Be$ to which one neutron is loosely bound, and {$^{6}\rm He$, described as an $\alpha$ core and two neutrons in the halo.}

Being very short lived, halo nuclei are often studied through reactions \cite{T96}. Breakup,  which corresponds to the dissociation of the halo from the core, is of particular interest since its cross section is 
high thanks to the fragile nature of the halo structure.
When performed on a heavy target, the reaction is dominated by the Coulomb interaction and the cross sections are large, which provides a clean probe of the projectile structure.
In particular it is often used to infer the E1 strength from the ground state to the continuum \cite{AN13}.
Coulomb breakup is thus also used as an indirect way to infer radiative-capture cross sections of astrophysical interest \cite{BBR86,BHT03}.

On light targets, breakup reactions are dominated by the nuclear force. 
Although the cross sections are {lower than on heavy targets, they carry other structure information, such as the energy and width of single-particle resonant states in the} continuum of the exotic projectile~\cite{Fetal04,CGB04}. 
When the halo neutron is not measured in coincidence with the core, the reaction is called one-neutron removal or knockout.
These reactions have been extensively used to study exotic nuclei since their statistics are significantly higher because they do not require the detection of the halo neutrons \cite{Aetal00,Sau01,Setal04}.
The corresponding cross sections have two contributions, one from the process---often referred to as the diffractive breakup---in which both the core and the halo neutrons survive the collision, and one from what is called stripping, i.e. channels in which the neutron is absorbed by the target.

Various studies have shown the significant role played by dynamical effects in both Coulomb- and nuclear-dominated breakup reactions and how they can affect the analysis of the data for both nuclear-structure and astrophysical applications \cite{TNT01,EBS05,CB05,SN08,Fetal12,MLG20,SMO20}.
The dynamics of the diffractive breakup is well treated by state-of-the-art models such as the continuum-discretized coupled channel method (CDCC) 
\cite{Kam86,TNT01,YOMM12},  time-dependent approach \cite{KYS94,EBB95,TW99,CBM03c} and eikonal-based models such as the eikonal-CDCC (E-CDCC) \cite{Oetal03,Oetal06} and the Dynamical Eikonal Approximation (DEA) \cite{BCG05,GBC06}.
The stripping contribution to the knockout reaction is most often analyzed within the Hussein-McVoy formalism, which is built upon the usual eikonal approximation, {including} an adiabatic treatment of the projectile structure \cite{G59,HM85,HBE96,HT03}.
In this framework, the projectile dynamics is therefore neglected~\cite{Tos02}.
A full dynamical description of the stripping process is thus needed to improve the analysis of knockout reactions and the quality of the nuclear-structure information inferred from these measurements.

With that goal in mind, the Kyushu-Osaka group has recently developed the eikonal reaction theory (ERT) \cite{YOM11}, which extends E-CDCC to stripping observables.
The underlying idea of the ERT is that thanks to its short range, the nuclear part of the projectile-target interaction, and in particular that between the halo neutron and the target, can be treated adiabatically, whereas the projectile dynamics cannot be ignored when treating the infinitely-ranged Coulomb force between the core and the target.
In this way, the neutron-target interaction can be accounted for through the usual eikonal $S$-matrix and the stripping contribution to the cross sections can be computed with the Hussein-McVoy formalism.
The ERT therefore enables the inclusion of the collision dynamics within the description of the stripping channel where it matters most \emph{a priori}, viz. in the treatment of the core-target interaction, while keeping the simplicity and elegance of the Hussein and McVoy modeling of the absorption of the valence neutron by the target.
In Ref.~\cite{YOM11}, Yahiro, Ogata and Minomo have used the ERT to study 
the one-neutron knockout  of $^{31}\rm Ne$ on $^{12}\rm C$ and 
$^{208}\rm Pb$ targets at about 230~MeV/nucleon.
They have then extended this method to two-neutrons knockout in Ref.~\cite{Metal14}, where they have analyzed the collision of $^6\rm He$ 
on the same targets at 240~MeV/nucleon and on $^{28}\rm Si$ at 52~MeV/nucleon.
For all these systems, they have compared integrated cross sections predicted by this ERT  to the ones obtained with the full E-CDCC calculations.
The uncertainty of this approximation on the total {diffractive-breakup} cross sections has been estimated to be within about 5\%~\cite{YOM11,Metal14}.

Because  these results are so encouraging, we apply in this article for the first time the ERT approach to the DEA, an eikonal-based model of reaction similar to E-CDCC \cite{FOC14}.
Yet, we go beyond the original idea of Yahiro, Ogata and Minomo \cite{YOM11} and analyze various ways to decompose the evolution operator and compare these results to {fully} dynamical DEA calculations of the diffractive breakup of $^{11}$Be on both light ($^{12}$C) and heavy ($^{208}$Pb) targets.
We also study how the approximation made within the ERT affects more differential observables, such as energy and parallel-momentum distributions.
We focus on the sole diffractive breakup because it is currently the only 
one that can be compared to accurate reaction calculations, in which dynamical effects are properly taken into account.

In this article, we proceed in the following way.
After presenting in Sec.~\ref{Sec2} the theoretical framework, we study in Sec.~\ref{Sec3} the validity of the ERT applied to the DEA for both nuclear- and Coulomb-dominated  reactions. 
We first verify that the ERT applied to the DEA leads to similar results for the integrated cross sections as the ones obtained with E-CDCC in Refs.~\cite{YOM11,Metal14}.
We then extend our analysis to differential cross sections, viz., the energy and parallel-momentum distributions, for which the ERT  has not been applied so far. 
Our conclusions are summarized in \Sec{Conclusions}.

\section{Reaction theory}\label{Sec2}
\subsection{Three-body model of reaction}
\begin{figure}
		\centering
		{\includegraphics[width=\linewidth]{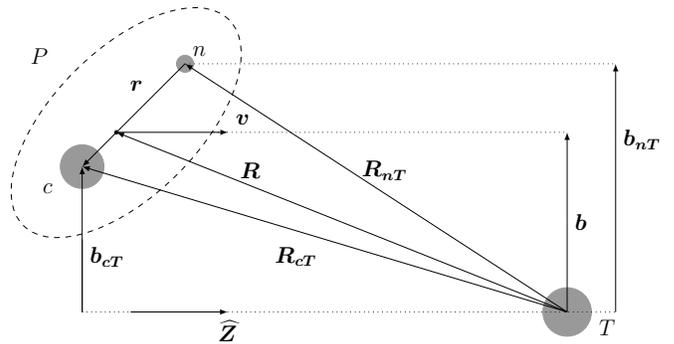}}
		\caption{\label{Fig3BodyCoordinates} Three-body model of the reaction and Jacobi set of coordinates. The projectile $P$, described as the $c$-$n$ two-body system, collides with the structureless target $T$. The $Z$ axis is chosen along the incoming beam. 
}
	\end{figure}

As usual, we base our description of the reaction on a few-body model \cite{BC12}.
The one-neutron halo nucleus projectile $P$ is described as a two-body system: a neutron $n$ loosely bound to a core $c$.
It impinges on the target $T$, which is assumed to be structureless.
This three-body model of the reaction and the coordinate system used below are illustrated in Fig.~\ref{Fig3BodyCoordinates}.

The internal structure of the projectile is described by the effective single-particle Hamiltonian
	\begin{equation}
h_{cn} = \frac{p^2 }{2 \mu_{cn}}  + V_{cn}(r).\label{eq1}
\end{equation}
where $\ve{p}$ and $\ve{r}$ are, respectively, the $c$-$n$ relative momentum
and coordinate, $\mu_{cn}$ is the $c$-$n$ reduced mass and $V_{cn}$ is an 
effective potential simulating the $c$-$n$ interaction. 

To account for the possible excitation of the target and other open channels, the interaction between the projectile fragments ($c$ and $n$) with the target are simulated by optical potentials ($V_{cT}$ and $V_{nT}$, respectively) chosen in the literature.
The resulting three-body model of the collision is described by the \Sch equation
\begin{eqnarray}
\left[\frac{P^2}{2\mu}+h_{cn}+V_{cT}+V_{nT}\right]\Psi(\ve{R},\ve{r})=E_{\rm tot}\ \Psi(\ve{R},\ve{r}), \label{eq2}
\end{eqnarray}
where $\ve{P}$ and $\ve R\equiv (\ve{b},Z)$ are respectively the $P$-$T$
relative momentum and coordinate (see Fig.~\ref{Fig3BodyCoordinates}), $\mu$ is the $P$-$T$ reduced mass, and $E_{\rm tot}$ is the total energy of 
the system.
This equation is solved with the initial condition that the projectile is 
in its ground state $\phi_{0}$ of energy $E_0$ and is impinging on the target 
\begin{equation}
	\Psi(\ve{R},\ve{r})\flim{Z}{- \infty}\exp(iKZ+\cdots)\ \phi_{0} (\ve{r})\label{eq3},
\end{equation}
where the $Z$ axis has been chosen along the incoming-beam direction, and 
the $P$-$T$ relative wave number $K$ is related to the total energy by energy conservation $E_{\rm tot}= E_{0}+ \frac{\hbar^2 K^2}{2\mu}$.
The	‘‘$\cdots$’’ in the exponential indicates that long-range interactions distort the plane wave even at large distances.

\subsection{Dynamical eikonal approximation}
The eikonal approximation reflects the fact that at high enough energy, the three-body wave function does not differ much from the initial plane wave.
This plane wave is hence factorized out of the wave function \cite{G59}
\begin{equation}
\Psi(\ve{R},\ve{r}) =\exp[iKZ]\ \widehat{\Psi}(\ve{R},\ve{r})\label{eq4}.
\end{equation}
Because that plane wave contains most of the $\ve{R}$-dependence of the three-body wave function $\Psi$, the new wave function $\widehat\Psi$ varies smoothly with the $P$-$T$ coordinate.
Accordingly its second-order derivative in $\ve{R}$ can be neglected compared to its first-order derivative.
This approximation simplifies the three-body \Sch \Eq{eq2} into the DEA equation~\cite{BCG05,GBC06,BC12}
\begin{eqnarray}
	i\hbar v \pderiv{}{Z} \widehat{\Psi}(\ve{R},\ve{r})=[h_{cn}-E_0+V_{cT}+V_{nT}]\widehat{\Psi}(\ve{R},\ve{r}),\label{eqDEA}
\end{eqnarray}	
where $v=\hbar K/\mu$ is the $P$-$T$ initial velocity.
Following \Eq{eq3}, the DEA \Eq{eqDEA} has to be solved with the condition $\widehat\Psi(\ve{R},\ve{r})\flim{Z}{-\infty}\phi_{0} (\ve{r})$.
As shown in Ref.~\cite{FOC14}, this description of the reaction is equivalent to the E-CDCC one \cite{Oetal03} for intermediate beam energies.

In the DEA, Eq.~\eqref{eqDEA} is solved through a numerical evolution calculation of the wave function along $Z$ for each $\ve b$ \cite{BCG05,GBC06}.
By using the unitary transformation 
 \begin{equation}
 \widetilde{\Psi}(\ve{R},\ve{r})=\exp[{i(h_{cn}-E_0) \frac{Z}{\hbar v}}]\widehat{\Psi}(\ve{R},\ve{r})\label{eq5},
 \end{equation}
the DEA \Eq{eqDEA} can be expressed as \cite{YOM11,Metal14,Hetal11,HB20}
\begin{eqnarray}
	i\hbar v \pderiv{}{Z} \widetilde{\Psi}(\ve{R},\ve{r})=
\hat{O}^\dagger[V_{cT}+V_{nT}]\hat{O}\widetilde{\Psi}(\ve{R},\ve{r})\label{eq6},
\end{eqnarray}
with the operator $\hat{O}$ defined as
\begin{eqnarray}
\hat{O}=\exp[{-ih_{cn}\frac{Z}{\hbar v}}]\label{eq7}.
\end{eqnarray}	
Formally, the  $S$-matrix operator of this system reads \cite{YOM11,Metal14,Hetal11}
\begin{equation}
\hat{S}=\exp\left\{-\frac{i}{\hbar v}\mathcal{P}\int_{-\infty}^{+\infty} \hat{O}^\dagger[V_{cT}+V_{nT}]\hat{O}\, dZ\right\}\label{eq8}
\end{equation}
where $\mathcal{P}$ is the path ordering operator describing the multistep scattering processes.

\subsection{Eikonal reaction theory}\label{ERT}

At intermediate or higher energies, the adiabatic---or sudden---approximation can be {in principle} safely made in breakup calculations for the treatment of short-range nuclear interactions.
This approximation sees the coordinate of the projectile as frozen during 
the collision and  holds only if the collision {time} is brief enough.
Consequently, it is not compatible with the infinite range of the  Coulomb interaction~\cite{MBB03,CBS08,HB20}.
The idea of the ERT is to exploit this idea and treat the purely nuclear $n$-$T$ interaction
at the adiabatic approximation, while keeping the dynamical treatment of the full $c$-$T$ potential \cite{YOM11,Metal14}.
{This is equivalent to assume $[V_{nT},\hat{O}]=0$ in \Eq{eq6}, which then reads}
\begin{eqnarray}
i \hbar v\pderiv{}{Z} \widetilde{\Psi}(\ve{R},\ve{r})=
[V_{nT}+\hat{O}^\dagger V_{cT}\hat{O}]\widetilde{\Psi}(\ve{R},\ve{r}).\label{eq9}
\end{eqnarray}	
 The ERT hence decouples the action of each interaction, leading to the factorization of the $S$-matrix operator~\eqref{eq8}
\begin{eqnarray}
\hat{S}^{\rm ERT}=S_n\,\exp[{-\frac{i}{\hbar v}\mathcal{P}\int_{-\infty}^{+\infty} \hat{O}^\dagger V_{cT}\hat{O}\,dZ}]\label{eq10}
\end{eqnarray}	
with $S_n=\exp[{-\frac{i}{\hbar v}\int_{-\infty}^{+\infty} V_{nT}\,dZ}]$ the usual $n$-$T$ eikonal $S$-matrix. 
This factorization of the $S$-matrix operator enables us to distinguish clearly the action of each interaction and allows the calculation of the stripping cross sections within the Hussein-McVoy formalism \cite{YOM11,Metal14,Hetal11}.
The main advantage of the ERT compared to the usual eikonal approximation 
is that the dynamical effects due to the $c$-$T$ field are accounted for.

If $V_{nT}$ commutes with the evolution operator $\hat{O}$, the effect of 
the $n$-$T$ $S$-matrix can be placed at any position $Z$ during the evolution calculation.
A more general factorization than the original ERT \eq{eq10} can thus be suggested
\begin{eqnarray}
\hat{S}^{{\rm ERT}^{(n)}(Z)}&=&\exp[{-\frac{i}{\hbar v}\mathcal{P}\int_{-\infty}^{Z} \hat{O}^\dagger V_{cT}\hat{O}\,dZ'}]\,S_n\nonumber \\
& &\times\exp[{-\frac{i}{\hbar v}\mathcal{P}\int_{Z}^{+\infty} \hat{O}^\dagger V_{cT}\hat{O}\,dZ'}].\label{eq10a}
\end{eqnarray}	
This expression {corresponds to a dynamical calculation with the $c$-$T$ interaction  performed up to $\ve{R}=\ve{b}+Z\ve{\hat Z}$}, at which point the full influence of $V_{nT}$ on the projectile is accounted for at 
the adiabatic approximation, before continuing the evolution calculation until $Z\rightarrow+\infty$.

Since the adiabatic approximation is in principle valid for the nuclear part of the $c$-$T$ interaction $V_{cT}^{(N)}$ as well, we also consider the following factorization
\begin{eqnarray}
\hat{S}^{{\rm ERT}^{(c)}(Z)}&=&\exp\left\{{-\frac{i}{\hbar v}\mathcal{P}\int_{-\infty}^{Z} \hat{O}^\dagger[ V_{nT}+V_{cT}^{(C)}]\hat{O}\,dZ'}\right\}\,S_c^{(N)}\nonumber\\
&\times&\exp\left\{{-\frac{i}{\hbar v}\mathcal{P}\int_{Z}^{+\infty} \hat{O}^\dagger[ V_{nT}+V_{cT}^{(C)}]\hat{O}\,dZ'}\right\},\label{eq11}
\end{eqnarray}	
with $S_c^{(N)}=\exp\left[-\frac{i}{\hbar v}\int_{-\infty}^{+\infty}V_{cT}^{(N)}\,dZ\right]$ the usual $c$-$T$ eikonal $S$-matrix.
The Coulomb part of the $c$-$T$ interaction $V_{cT}^{(C)}$ cannot be treated at the adiabatic approximation and must be included in the dynamical evolution.

We study the approximations \eq{eq10}, \eqref{eq10a} and \eqref{eq11} first for  integrated cross section as in Ref.~\cite{YOM11}.
We extend that study to the choice of the particular value of $Z$ at which the eikonal $S$-matrix is included.
We then look more closely at the quality of the ERT and its extensions to 
describe the energy and  parallel-momentum distributions for both nuclear- and Coulomb-dominated breakup reactions of a one-neutron nucleus.

\section{Results}\label{Sec3}
\subsection{Two-body interactions}\label{Sec3a}

To assess the validity of the ERT and its variants described in \Sec{ERT}, we systematically compare the ERT approximation to reliable DEA calculations of the breakup of $^{11}$Be with $^{12}$C at 67~MeV/nucleon and with $^{208}$Pb at 69~MeV/nucleon.
These reactions have both been measured at RIKEN \cite{Fetal04} and are well described at the DEA \cite{GBC06,CPH18}.

The $^{11}$Be nucleus has only two bound states: a $1/2^+$ ground state with a one-neutron separation energy of 504~keV and a $1/2^-$ excited state bound by 184~keV.
Within the single-particle approximation used here, these states are described as an inert $^{10}$Be core in its $0^+$ ground state with a valence 
neutron in the $1s_{1/2}$ and $0p_{1/2}$ orbitals, respectively.
We follow Refs.~\cite{CPH18,HC19} and simulate the $^{10}$Be-$n$ interaction through a Halo-EFT expansion at next-to-leading-order (NLO)~\cite{BHvk02,BHvk03} (see Ref.~\cite{HJP17} for a recent review).
In this approach, we consider a Gaussian shaped effective potential in the $s_{1/2}$ and $p_{1/2}$ waves to generate $^{11}$Be bound states.
These potentials are fit to the experimental binding energies of these states as well as the asymptotic normalization constants predicted by the \textit{ab initio} calculations of Calci \etal~\cite{Cetal16}.
These potentials also reproduce the low-energy behavior of the $s_{1/2}$ and $p_{1/2}$ \emph{ab initio} phaseshifts.
The effective interaction is chosen to be nil in the $p_{3/2}$ partial wave to match the nil \emph{ab initio} phaseshift in that partial wave at low energy.
We use the parametrization (13) of Ref.~\cite{CPH18} and the depths listed 
in its Tables I and II corresponding to $\sigma=1.2$~fm.
Because that state plays a significant role in nuclear breakup \cite{Fetal04,CGB04}, we also include the $5/2^+$ resonance at 1.274~MeV in the $^{10}$Be-$n$ continuum.
For this we go beyond NLO, and add an effective interaction in the $d_{5/2}$ partial wave.
We  take the parameters corresponding to $\sigma=1.2$~fm which are displayed in Table IV of Ref.~\cite{CPH18}.
This description of $^{11}$Be leads to excellent descriptions of breakup \cite{CPH18,MC19}, transfer \cite{YC18}, and knockout \cite{PhDHebborn,HC21} reactions.

The $^{10}$Be-$T$ and  $n$-$T$ interactions are simulated through the same optical potentials as in Ref.~\cite{CBS08}.
The DEA calculations are performed with the same numerical inputs as in Ref.~\cite{CBS08}. The total breakup cross sections are computed by numerically integrating the energy distribution up to $E=33$~MeV in the $^{10}$Be-$n$ continuum.

\subsection{Diffractive-breakup observables}\label{Sec3b}

\subsubsection{Total diffractive-breakup cross sections}\label{Sec3b1}

\begin{table}
	\begin{tabular}[c]{c|c|c|c|c}\hline \hline
		Target& DEA &ERT&ERT$^{(n)}(Z=0)$&ERT$^{(c)}(Z=0)$\\\hline
		$^{12}$C& 144 mb&138 mb&138 mb& 146 mb\\
		$^{208}$Pb&1890 mb&1925 mb&1915 mb&1895 mb\\
		\hline \hline
	\end{tabular}
	\caption{Total diffractive-breakup cross sections obtained with the DEA, {the original  ERT \Eq{eq10}, the extension ERT$^{(n)}$ \Eq{eq10a} with $Z=0$}, and the ERT$^{(c)}$ \Eq{eq11} with $Z=0$, for  $^{11}$Be impinging on $^{12}$C at 67~MeV/nucleon and $^{208}$Pb at 69~MeV/nucleon. }
	\label{t1}
\end{table}

In its original derivation, the ERT approximation was applied to E-CDCC, and its predictions were compared to these dynamical calculations for total diffractive-breakup cross sections \cite{YOM11,Metal14}.
Small uncertainties, within about 5\%, have been observed for different reactions.
In this first step, we perform a similar analysis in our implementation of the ERT within the DEA \Eq{eq10}; we also extend this test to the ERT formul\ae~\eq{eq10a} and \eq{eq11} developed in \Sec{ERT}.
Our results are displayed in Table~\ref{t1} for the breakup of $^{11}$Be on $^{12}$C at 67~MeV/nucleon and on $^{208}$Pb at 69~MeV/nucleon.

\begin{figure*}[ht]
	\centering
	{\includegraphics[width=0.48\linewidth]{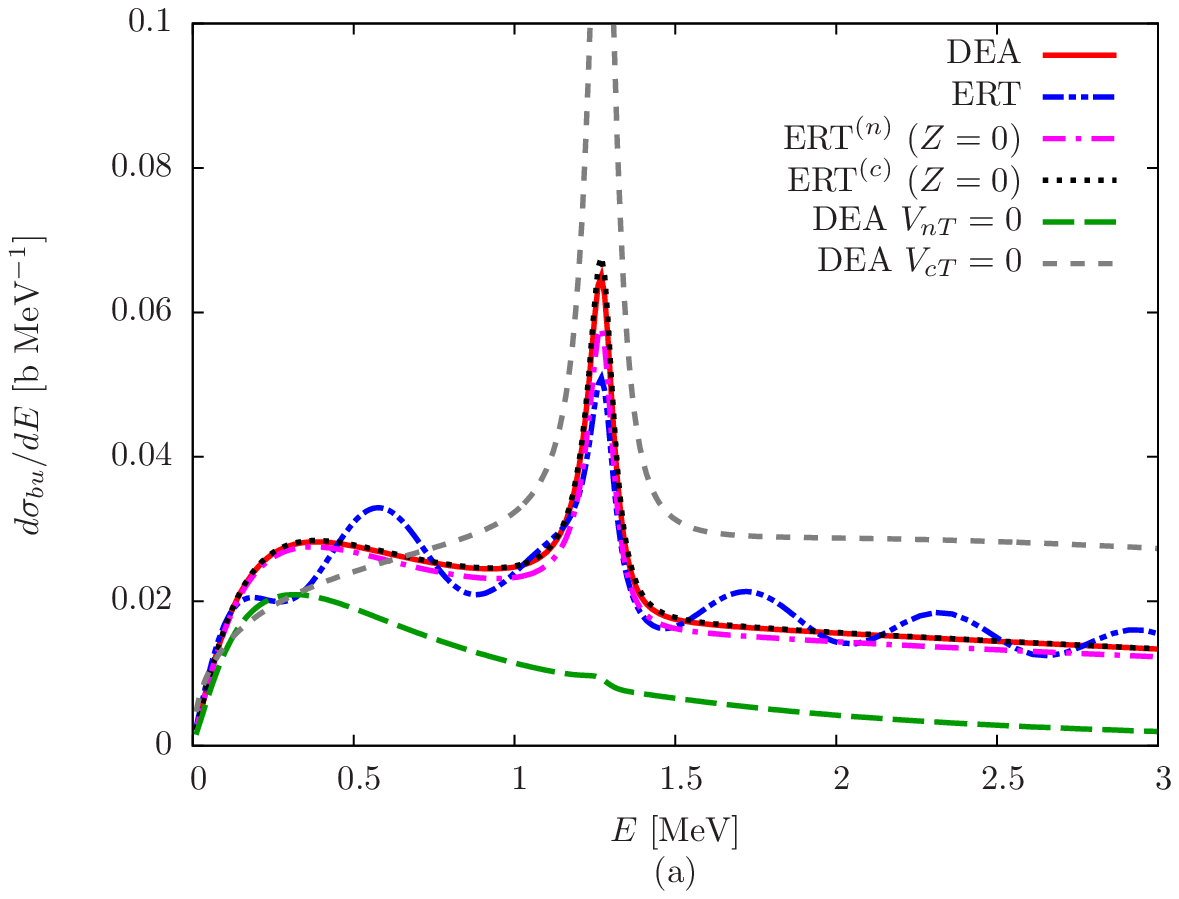}}\hspace{0.4cm}
	{\includegraphics[width=0.48\linewidth]{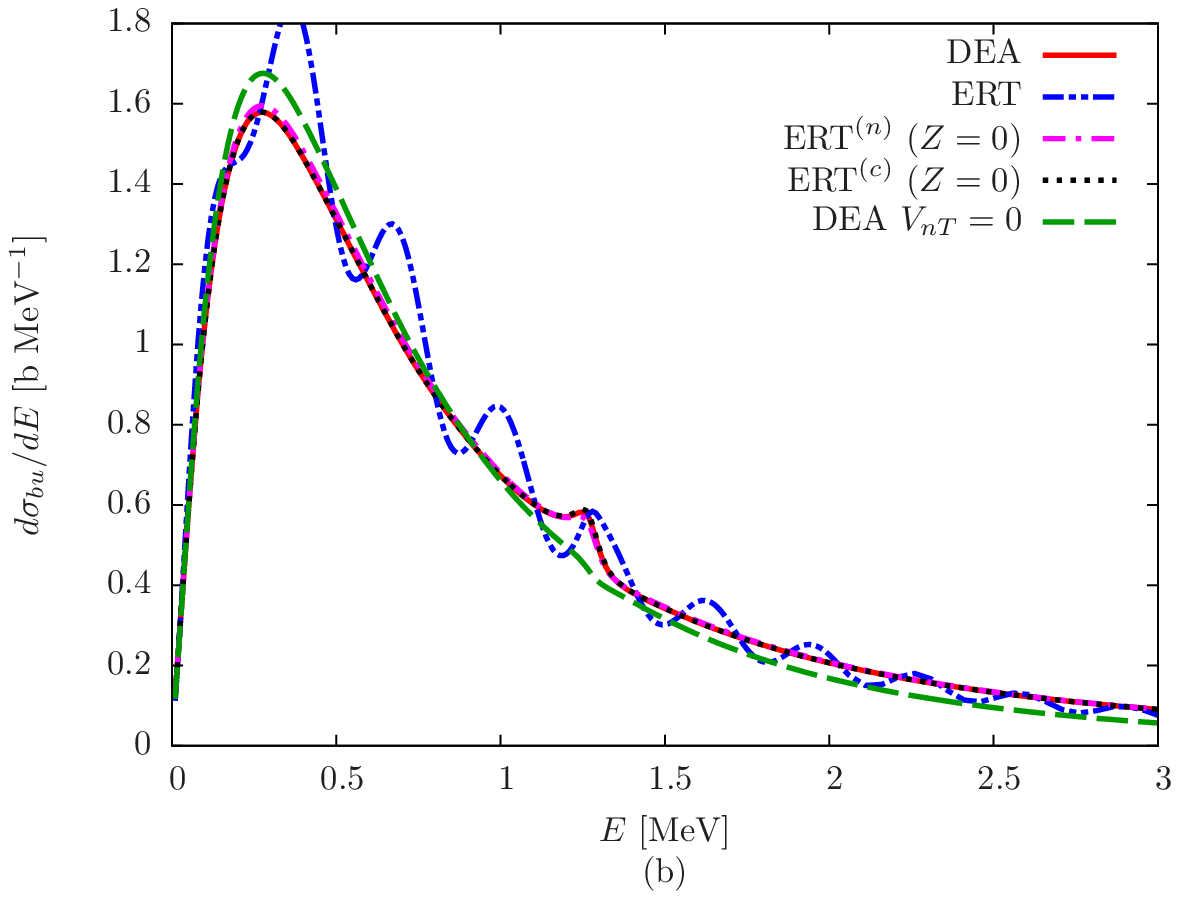}}
	\caption{Breakup cross section of $^{11}$Be as a function of the $^{10}$Be-$n$ relative energy $E$ on (a) $^{12}$C at 67~MeV/nucleon and 
(b)~$^{208}$Pb at 69~MeV/nucleon.
	The DEA reference calculation is compared to different implementations of the ERT: the original ERT \Eq{eq10}, ERT$^{(n)}$ \Eq{eq10a} with $Z=0$, and ERT$^{(c)}$ \Eq{eq11} with $Z=0$.
	The roles of $V_{cT}$ and $V_{nT}$ are also tested.  
	}\label{f2}
\end{figure*}

The first implementation of the ERT within the DEA~{\eq{eq10} is faithful  to
the original idea of Yahiro, Ogata and Minomo by treating the $n$-$T$ adiabatically 
after the dynamical treatment of the $c$-$T$ interaction.
Here we also consider a calculation ERT$^{(n)}$ \eq{eq10a}} with $Z=0$, i.e. in which $S_n$ is placed at the distance of closest approach between the projectile and the 
target, which, in a semi-classical description of the reaction, makes more sense.
On the carbon target, the value of $Z$ has no influence on the total cross section; other tests, including {ERT$^{(n)}$ \eq{eq10a} with} $Z\rightarrow-\infty$, have confirmed this independence.
These results are very close to the DEA calculation (the ERT estimates are about 5\% lower than the DEA).
On Pb, the value of $Z$ affects the results a bit more, but the difference between {\Eq{eq10} and \Eq{eq10a} with $Z=0$} remains within a negligible 0.5\%.
In this case, the ERT estimates of the cross section slightly surpass the 
reference calculation (1\% difference with the DEA).

In our implementation {ERT$^{(c)}$} \eq{eq11}, it is the nuclear part of the $c$-$T$ interaction that is treated adiabatically.
Interestingly, this performs even better than the first implementation.
The cross sections are almost exactly the same as the ones obtained from the full DEA on both targets.

These tests confirm that for integrated breakup cross sections, the nuclear part of the $P$-$T$ interaction can be safely treated adiabatically. 
As observed initially in Refs.~\cite{YOM11,Metal14,Hetal11}, the differences with the full 
dynamical calculations remain within about 5\%, which is significantly smaller than the uncertainty related to the choice of the optical potentials, see, 
e.g., Ref.~\cite{CGB04}.
In the following two sections, we extend this study to differential cross 
sections that are often measured in the study of halo nuclei: the energy distribution (\Sec{Sec3b2}) and the parallel-momentum distribution (\Sec{Sec3b3}).

\subsubsection{Energy distributions}\label{Sec3b2}

The cross sections for the diffractive breakup of $^{11}$Be are displayed in \Fig{f2} as a function of the relative energy $E$ between the $^{10}$Be core and the halo neutron $n$ after dissociation on (a) $^{12}$C at 67~MeV/nucleon and (b)~$^{208}$Pb at 69~MeV/nucleon.
In addition to the DEA reference calculations (solid red lines), the results of the three ERT implementations considered in the previous section are also shown.
The original ERT of Ref.~\cite{YOM11}, i.e. {using \Eq{eq10}}, is shown in dash-dotted-dotted blue lines.
The ERT implementation with the adiabatic treatment of $V_{nT}$ set at the $P$-$T$ distance of closest approach, i.e. corresponding to \Eq{eq10a} with $Z=0$, is displayed in dash-dotted magenta lines.
When the nuclear part of the $c$-$T$ interaction is treated adiabatically 
[\Eq{eq11} with $Z=0$], we obtain the dotted black lines.

Prior to analyzing the different implementations of the ERT, let us mention that, once folded with the experimental energy resolution, the DEA calculations are in excellent agreement with the RIKEN data (see Ref.~\cite{CPH18}).
They thus correspond to very accurate descriptions of the reaction on both targets.

\begin{figure*}[ht]
	\centering
	{\includegraphics[width=0.48\linewidth]{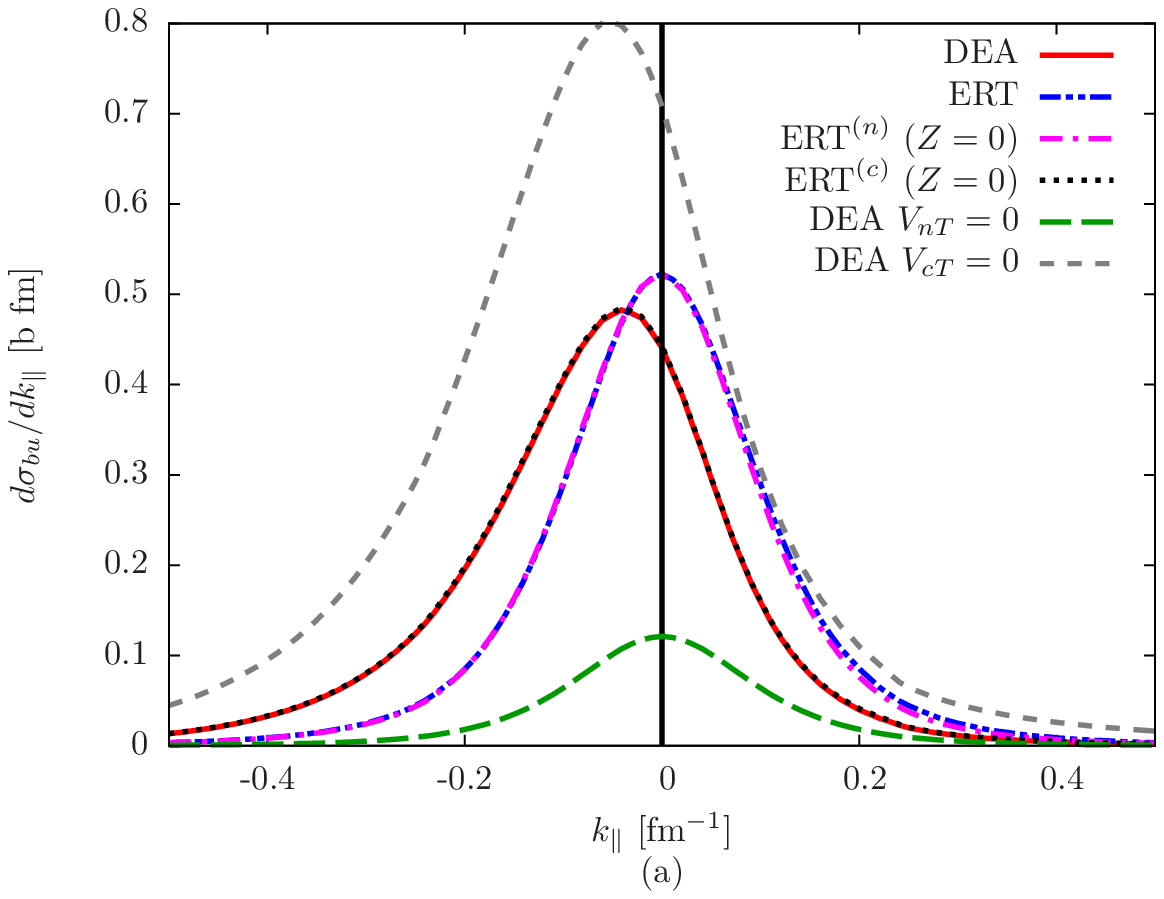}}\hspace{0.4cm}
	{\includegraphics[width=0.48\linewidth]{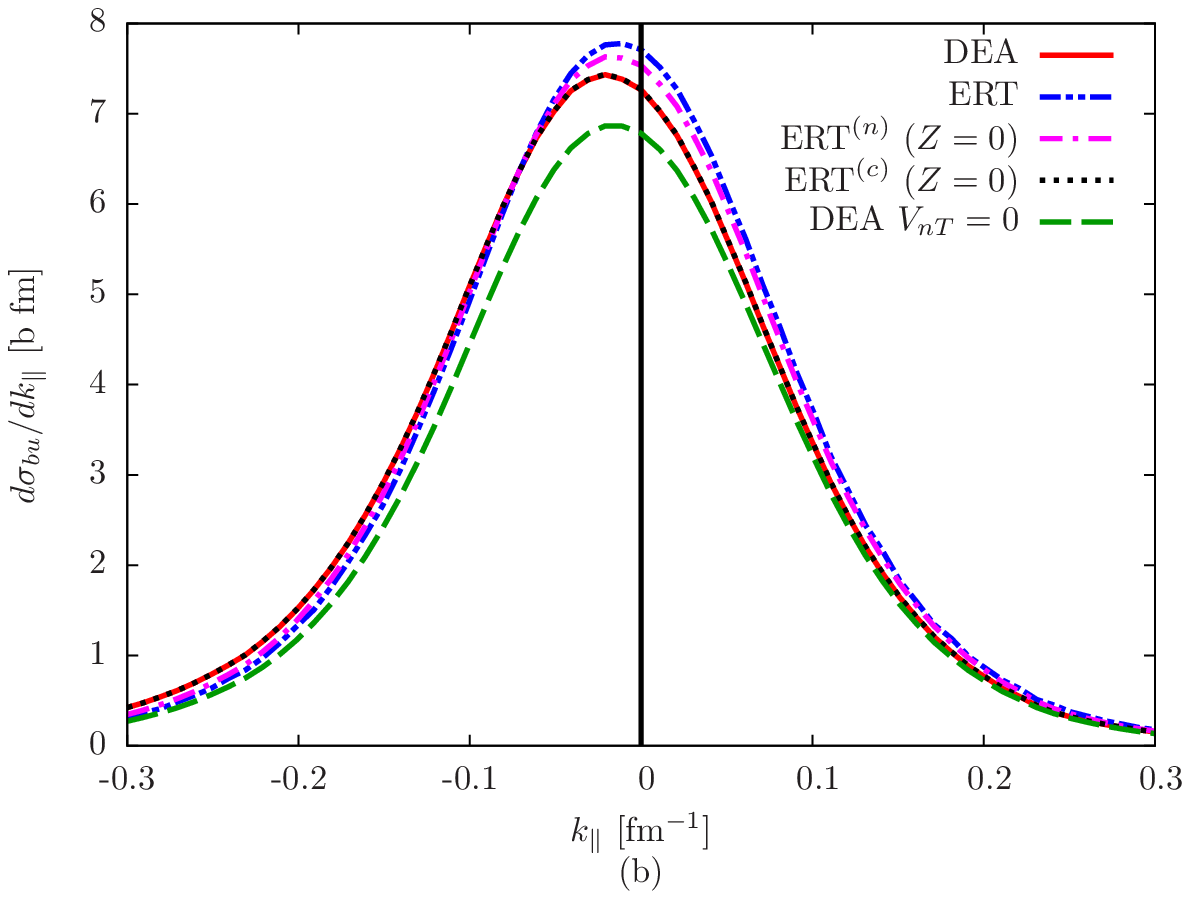}}	
	\caption{Cross sections as a function of the {$^{10}\rm Be$-$n$} parallel-momentum for the diffractive breakup of $^{11}\rm Be$ on (a) $^{12}\rm C$ at 67~MeV/nucleons and (b) $^{208}\rm Pb$ at 69~MeV/nucleons.  The results using various implementations of the ERT are compared to the reference DEA calculations. The roles of $V_{nT}$ and {$V_{cT}$} are also explored.
}\label{f3}
\end{figure*}

The original ERT results exhibit unphysical oscillations around the DEA cross sections.
Additional calculations have shown that these oscillations also appear if 
the  $n$-$T$ eikonal $S$-matrix is applied before the dynamical treatment 
of $V_{cT}$, hence {using \Eq{eq10a} with} $Z\rightarrow-\infty$, and that their period varies with the initial and final values of $Z$ in the numerical estimate of the dynamical calculation. 
This adiabatic treatment of $V_{nT}$ therefore interferes with the dynamical treatment of $V_{cT}$, which, at the first order of the perturbation theory, includes a phase $\exp\left[i (E-E_0) Z/v\right]$ \cite{CBS08,HB20}.
Interestingly, when the effect of the $n$-$T$ interaction $S_n$ is placed 
within the evolution calculation at the distance of closest approach between the projectile center of mass and the target, i.e. at $Z=0$ in \Eq{eq10a}, the oscillations completely vanish (see the dash-dotted magenta lines in \Fig{f2}) and are in very good agreement with the {fully} dynamical calculations.
In a semiclassical interpretation of this eikonal formalism, this choice makes sense: if $V_{nT}$ can be treated adiabatically due to its short range, its effect on the projectile structure will be strongly localized where the projectile is closest to the target.
The presence of these oscillations is not observed in the results shown in Table~\ref{t1} because these total breakup cross sections are 
obtained after integration over the continuum energy $E$.

The adiabatic approximation applied to the nuclear part of the $^{10}\rm Be$-$T$ interaction~\eqref{eq11} performs extremely well for both targets 
(see the dotted black lines in \Fig{f2}).
As already observed in Table~\ref{t1}, they are nearly superimposed to the {fully} dynamical calculations.
This shows that the adiabatic approximation is even more justified in the 
treatment of the short-ranged $c$-$T$ interaction.
To better grasp this difference, we repeat the DEA calculations without the $n$-$T$ interaction (long-dashed green lines in \Fig{f2}) and, only for the light target, without the {full $c$-$T$ interaction [short-dashed grey line in \Fig{f2}].

On $^{12}$C, we observe that the $n$-$T$ interaction is the driving force 
of the breakup.
Without it, the breakup cross section looks similar to a pure Coulomb calculation, see Fig.~3 of Ref.~\cite{CGB04}.
The breakup at high energy $E$ and the population of the $d_{5/2}$ resonance takes place mostly thanks to that interaction, i.e. through the scattering of the halo neutron by the target, while the core moves unperturbed, as a spectator.
When $V_{cT}$  is set to zero, most of the breakup cross section lies above the DEA calculation.
This indicates that, {except at low $E$, where the Coulomb part of $V_{cT}$ plays a role, the major influence of that interaction is to remove flux from the reaction channel through nuclear absorption.}
Since that effect is well captured by the eikonal phase $S_c^{(N)}$, it explains why the ERT implementation \eq{eq11} works so well.

On the Pb target, the breakup process is Coulomb dominated and $V_{nT}$ has only a minor effect on the breakup cross section.
We can nevertheless note that its presence, albeit small, is visible at all energy and is necessary to populate the $d_{5/2}$ resonance.
{Because the breakup is dominated by $V_{cT}^{(C)}$, neglecting the $c$-$T$ interaction on such a heavy target would be meaningless. 
We have therefore not represented that calculation in \Fig{f2}(b).}

This series of tests shows that, but for unphysical oscillations, the ERT 
fairly well reproduces dynamical calculations of the diffractive-breakup cross section for one-neutron halo nuclei expressed as a function of the core-neutron relative energy after dissociation, and this on both light and heavy targets.
In the following section, we study in more detail the influence of this choice upon the parallel-momentum distribution, which is the observable usually measured in knockout reactions \cite{Aetal00,Sau01,Setal04,HT03}, and which is highly sensitive to dynamical effects \cite{CBS08,Fetal12,Tos02}.

\subsubsection{Parallel-momentum observables}\label{Sec3b3}

Fig.~\ref{f3} shows the cross section for the diffractive breakup of $^{11}$Be as a function of the component parallel to the beam axis $k_{\parallel}$ of the {$^{10}$Be-$n$ relative wave vector.} 
As in the previous sections, the calculations are performed for (a) a $^{12}$C target at 67~MeV/nucleon and (b) a $^{208}$Pb target at 69~MeV/nucleon.

As observed in Refs.~\cite{Tos02,CBS08}, dynamical calculations (solid red lines) predict a slightly asymmetric parallel-momentum distribution, in agreement with what is observed experimentally \cite{Aetal00,Sau01,Setal04,Fetal12}.
This asymmetry appears for both targets, despite the very different reaction mechanisms.
The capability of a reaction model to describe this feature is central in 
the study of knockout reactions.

The original implementation of the ERT [\Eq{eq10}; dash-dotted-dotted blue lines] predicts smooth cross sections because the oscillations observed in \Fig{f2} cancel through the integration over energy involved in the calculation of that observable.
Unfortunately, it predicts a symmetric distribution on $^{12}$C, and on $^{208}$Pb its cross section does not exhibit as much asymmetry as the DEA 
one.
The implementation ERT$^{(n)}$ \Eq{eq10a} with  $Z=0$, does not correct for that flaw (dash-dotted magenta lines).
This suggests that this description of the reaction, despite its capability to reproduce the integrated breakup cross sections (see Table~\ref{t1}) and the energy distributions (see \Fig{f2}), misses a significant aspect of the reaction mechanism, especially on light targets, i.e. in nuclear-dominated collisions.
Accounting for $V_{nT}$ within the dynamical calculation, while treating the nuclear part of the $c$-$T$ interaction adiabatically [\Eq{eq11} with 
$Z=0$; dotted black lines], perfectly reproduces the DEA parallel-momentum distribution.

The results shown in \Fig{f3}(a) illustrate the key role played by the $n$-$T$ interaction in nuclear breakup and the necessity to include its dynamical treatment within the reaction model in order to correctly reproduce this reaction observable usually measured in knockout experiments.
When that interaction is switched off (long-dashed green line), the DEA cross section drops by a factor of five, and is purely symmetric.
On the contrary, when the  $c$-$T$ interaction is switched off, the dynamical cross section (short-dashed grey line) nearly doubles while remaining significantly asymmetric.
As already noted in the discussion on the energy distributions, the dominant contribution of $V_{cT}^{(N)}$ in the reaction process is the absorption from the diffractive-breakup channel, which is well described with the usual eikonal phase $S_c^{(N)}$ in \Eq{eq11}.
This is why this implementation of the ERT is so efficient in describing all three diffractive-breakup observables considered here.

The difference between these two interactions is a signature from the halo structure of the projectile.
The core, with its large mass compared to that of the valence neutron, follows roughly the eikonal trajectory of the center of mass of the projectile.
The nuclear part of its interaction with the target is thus mostly localized to short $P$-$T$ distances.
Because of the large extension of its wave function, the halo neutron can 
interact with the target even when the projectile center of mass is far from it.
The range of $P$-$T$ distances over which $V_{nT}$ influences the reaction is thus much larger than for the core.
Accordingly, the adiabatic treatment of that interaction is less valid than for the $c$-$T$ one.
Note that for a halo nucleus with a much lighter core, such as a deuteron, the adiabatic approximation would be much likely questionable for both $V_{nT}$ and $V_{cT}^{(N)}$.

The role of $V_{nT}$ in the diffractive breakup on $^{208}$Pb is much more modest.
Without it, the dynamical cross section is reduced by only 10\% and the asymmetry of the DEA cross section is only slightly underestimated [see long-dashed green line in \Fig{f3}(b)].
The adiabatic treatment of the $n$-$T$ interaction suggested in the original ERT partially corrects for these differences.

\section{Conclusions}\label{Conclusions}

Knockout reactions are often used to study the structure of halo nuclei \cite{Aetal00,Sau01,Setal04}.
Their {application has also been extended as a spectroscopic tool to explore the single-particle structure of other} short-lived nuclei \cite{HT03,Tos02}.
These experimental studies are usually coupled to an eikonal model of the 
reaction, which includes an adiabatic treatment of the reaction dynamics \cite{G59,HM85,HBE96,HT03}.
This model predicts an exactly symmetric parallel-momentum distribution of the core of the nucleus after removal of the valence nucleon, whereas experimental data exhibit a clear asymmetric cross section \cite{Aetal00,Tos02,Fetal12}.
This flaw has been explained as the lack of dynamics in the eikonal description of the reaction \cite{Tos02}.
To correct it, Yahiro, Ogata and Minomo have introduced the ERT \cite{YOM11}, which enables them to account for the reaction dynamics in the treatment of the $c$-$T$ interaction while describing the effect of $V_{nT}$ adiabatically.
Within this ERT, only integrated cross sections have been studied so far \cite{YOM11,Metal14,Hetal11}

We have analyzed in detail the application of the ERT to the diffractive breakup of one-neutron halo nuclei.
In particular, we have extended the original idea of Ref.~\cite{YOM11} and explored its validity in the calculation of differential cross sections 
expressed as a function of the $c$-$n$ relative energy and {momentum} after 
dissociation.
We have compared predictions of the ERT to accurate DEA calculations of the breakup of $^{11}$Be on $^{12}$C at 67~MeV/nucleon and $^{208}$Pb at 69~MeV/nucleon.

{For} the integrated diffractive-breakup cross sections, the ERT exhibits an error of about 5\% compared to the dynamical calculation, which is consistent with the uncertainty cited in Refs.~\cite{YOM11,Metal14}.
Unfortunately, the original implementation of the ERT produces unphysical 
oscillations in the energy distributions.
This can be avoided by placing the eikonal phase for the $n$-$T$ interaction $S_n$ in the middle of the evolution calculation, at the $P$-$T$ distance of closest approach.
This minor change produces a cross section in very good agreement with the DEA calculations.

{In the parallel-momentum distributions, however, the adiabatic treatment of $V_{nT}$ assumed in the original ERT does not produce a cross section in full agreement with the dynamical calculations.
Especially on the light target}, it lacks the asymmetry of the distribution observed in experimental data \cite{Aetal00,Tos02,Fetal12}.
This issue can be corrected only if the $n$-$T$ interaction is included in the dynamical evolution.
This complicates the original idea of Ref.~\cite{YOM11} to apply the ERT to include the missing dynamical effects of the usual eikonal approximation {in order} to analyze more reliably data from knockout experiments.
Interestingly, though, the nuclear part of $V_{cT}$ can be safely treated 
adiabatically, even on a heavy target, which simplifies the reaction model.
This difference between both interactions is related to the halo structure of the projectile.
The large extension of the wave function of the halo neutron makes its interaction with the target {less likely to take} place on a short distance, and hence that it can be treated at the adiabatic---or sudden---approximation.

In conclusion, this study confirms the hypothesis of the ERT that short-range nuclear interactions can be approximately treated adiabatically in a 
few-body model of breakup reactions.
Unfortunately, the original idea presented in Ref.~\cite{YOM11} misses part of the dynamics  of the reaction and therefore does not grasp the full 
reaction mechanism.
Accordingly, the one-neutron removal cross section computed with this approach are probably not as precise as first hoped.
Our detailed study has clearly shown that an accurate description of these observables requires a dynamical treatment of the neutron-target interaction.
An extension of the Hussein-McVoy formalism that includes this dynamics is needed to accurately describe knockout reactions in order to infer 
reliable structure information from experimental data.

\begin{acknowledgements}
	{C.~H. acknowledges the support of the Belgian Fund for Research Training in 
Industry and Agriculture (FRIA) and of the U.S. Department of Energy, Office of Science, Office of Nuclear Physics, under the FRIB Theory Alliance award DE-SC0013617.}
This project has received funding from the European Union’s Horizon 2020 research and innovation program under grant agreement No 654002, the Fonds de la 
Recherche Scientifique - FNRS under Grant Number 4.45.10.08,  the Deutsche Forschungsgemeinschaft Projekt-ID 279384907 -- SFB 1245 and Projekt-ID 204404729 -- SFB 1044, and the PRISMA+ (Precision Physics, Fundamental Interactions and Structure of Matter) Cluster of Excellence. 
P.~C. acknowledges the support of the State of Rhineland-Palatinate.
\end{acknowledgements}

\bibliographystyle{apsrev}
\bibliography{HC_ERT}

\begin{thebibliography}{51}
\expandafter\ifx\csname natexlab\endcsname\relax\def\natexlab#1{#1}\fi
\expandafter\ifx\csname bibnamefont\endcsname\relax
  \def\bibnamefont#1{#1}\fi
\expandafter\ifx\csname bibfnamefont\endcsname\relax
  \def\bibfnamefont#1{#1}\fi
\expandafter\ifx\csname citenamefont\endcsname\relax
  \def\citenamefont#1{#1}\fi
\expandafter\ifx\csname url\endcsname\relax
  \def\url#1{\texttt{#1}}\fi
\expandafter\ifx\csname urlprefix\endcsname\relax\def\urlprefix{URL }\fi
\providecommand{\bibinfo}[2]{#2}
\providecommand{\eprint}[2][]{\url{#2}}

\bibitem[{\citenamefont{Tanihata et~al.}(1985)\citenamefont{Tanihata, Hamagaki,
  Hashimoto, Shida, Yoshikawa, Sugimoto, Yamakawa, Kobayashi, and
  Takahashi}}]{Tetal85a}
\bibinfo{author}{\bibfnamefont{I.}~\bibnamefont{Tanihata}},
  \bibinfo{author}{\bibfnamefont{H.}~\bibnamefont{Hamagaki}},
  \bibinfo{author}{\bibfnamefont{O.}~\bibnamefont{Hashimoto}},
  \bibinfo{author}{\bibfnamefont{Y.}~\bibnamefont{Shida}},
  \bibinfo{author}{\bibfnamefont{N.}~\bibnamefont{Yoshikawa}},
  \bibinfo{author}{\bibfnamefont{K.}~\bibnamefont{Sugimoto}},
  \bibinfo{author}{\bibfnamefont{O.}~\bibnamefont{Yamakawa}},
  \bibinfo{author}{\bibfnamefont{T.}~\bibnamefont{Kobayashi}},
  \bibnamefont{and}
  \bibinfo{author}{\bibfnamefont{N.}~\bibnamefont{Takahashi}},
  \bibinfo{journal}{Phys. Rev. Lett.} \textbf{\bibinfo{volume}{55}},
  \bibinfo{pages}{2676} (\bibinfo{year}{1985}).

\bibitem[{\citenamefont{Tanihata}(1996)}]{T96}
\bibinfo{author}{\bibfnamefont{I.}~\bibnamefont{Tanihata}},
  \bibinfo{journal}{J.~Phys.~G} \textbf{\bibinfo{volume}{22}},
  \bibinfo{pages}{157} (\bibinfo{year}{1996}).

\bibitem[{\citenamefont{Hansen and Jonson}(1987)}]{HJ87}
\bibinfo{author}{\bibfnamefont{P.~G.} \bibnamefont{Hansen}} \bibnamefont{and}
  \bibinfo{author}{\bibfnamefont{B.}~\bibnamefont{Jonson}},
  \bibinfo{journal}{Europhys. Lett.} \textbf{\bibinfo{volume}{4}},
  \bibinfo{pages}{409} (\bibinfo{year}{1987}).

\bibitem[{\citenamefont{Aumann and Nakamura}(2013)}]{AN13}
\bibinfo{author}{\bibfnamefont{T.}~\bibnamefont{Aumann}} \bibnamefont{and}
  \bibinfo{author}{\bibfnamefont{T.}~\bibnamefont{Nakamura}},
  \bibinfo{journal}{Phys. Script.} \textbf{\bibinfo{volume}{T152}},
  \bibinfo{pages}{014012} (\bibinfo{year}{2013}).

\bibitem[{\citenamefont{Baur et~al.}(1986)\citenamefont{Baur, Bertulani, and
  Rebel}}]{BBR86}
\bibinfo{author}{\bibfnamefont{G.}~\bibnamefont{Baur}},
  \bibinfo{author}{\bibfnamefont{C.}~\bibnamefont{Bertulani}},
  \bibnamefont{and} \bibinfo{author}{\bibfnamefont{H.}~\bibnamefont{Rebel}},
  \bibinfo{journal}{Nucl. Phys. A} \textbf{\bibinfo{volume}{458}},
  \bibinfo{pages}{188} (\bibinfo{year}{1986}).

\bibitem[{\citenamefont{Baur et~al.}(2003)\citenamefont{Baur, Hencken, and
  Trautmann}}]{BHT03}
\bibinfo{author}{\bibfnamefont{G.}~\bibnamefont{Baur}},
  \bibinfo{author}{\bibfnamefont{K.}~\bibnamefont{Hencken}}, \bibnamefont{and}
  \bibinfo{author}{\bibfnamefont{D.}~\bibnamefont{Trautmann}},
  \bibinfo{journal}{Prog. Part. Nucl. Phys.} \textbf{\bibinfo{volume}{51}},
  \bibinfo{pages}{487} (\bibinfo{year}{2003}).

\bibitem[{\citenamefont{Fukuda et~al.}(2004)\citenamefont{Fukuda, Nakamura,
  Aoi, Imai, Ishihara, Kobayashi, Iwasaki, Kubo, Mengoni, Notani
  et~al.}}]{Fetal04}
\bibinfo{author}{\bibfnamefont{N.}~\bibnamefont{Fukuda}},
  \bibinfo{author}{\bibfnamefont{T.}~\bibnamefont{Nakamura}},
  \bibinfo{author}{\bibfnamefont{N.}~\bibnamefont{Aoi}},
  \bibinfo{author}{\bibfnamefont{N.}~\bibnamefont{Imai}},
  \bibinfo{author}{\bibfnamefont{M.}~\bibnamefont{Ishihara}},
  \bibinfo{author}{\bibfnamefont{T.}~\bibnamefont{Kobayashi}},
  \bibinfo{author}{\bibfnamefont{H.}~\bibnamefont{Iwasaki}},
  \bibinfo{author}{\bibfnamefont{T.}~\bibnamefont{Kubo}},
  \bibinfo{author}{\bibfnamefont{A.}~\bibnamefont{Mengoni}},
  \bibinfo{author}{\bibfnamefont{M.}~\bibnamefont{Notani}},
  \bibnamefont{et~al.}, \bibinfo{journal}{Phys. Rev. C}
  \textbf{\bibinfo{volume}{70}}, \bibinfo{pages}{054606}
  (\bibinfo{year}{2004}).

\bibitem[{\citenamefont{Capel et~al.}(2004)\citenamefont{Capel, Goldstein, and
  Baye}}]{CGB04}
\bibinfo{author}{\bibfnamefont{P.}~\bibnamefont{Capel}},
  \bibinfo{author}{\bibfnamefont{G.}~\bibnamefont{Goldstein}},
  \bibnamefont{and} \bibinfo{author}{\bibfnamefont{D.}~\bibnamefont{Baye}},
  \bibinfo{journal}{Phys. Rev. C} \textbf{\bibinfo{volume}{70}},
  \bibinfo{pages}{064605} (\bibinfo{year}{2004}).

\bibitem[{\citenamefont{Aumann et~al.}(2000)\citenamefont{Aumann, Navin,
  Balamuth, Bazin, Blank, Brown, Bush, Caggiano, Davids, Glasmacher
  et~al.}}]{Aetal00}
\bibinfo{author}{\bibfnamefont{T.}~\bibnamefont{Aumann}},
  \bibinfo{author}{\bibfnamefont{A.}~\bibnamefont{Navin}},
  \bibinfo{author}{\bibfnamefont{D.~P.} \bibnamefont{Balamuth}},
  \bibinfo{author}{\bibfnamefont{D.}~\bibnamefont{Bazin}},
  \bibinfo{author}{\bibfnamefont{B.}~\bibnamefont{Blank}},
  \bibinfo{author}{\bibfnamefont{B.~A.} \bibnamefont{Brown}},
  \bibinfo{author}{\bibfnamefont{J.~E.} \bibnamefont{Bush}},
  \bibinfo{author}{\bibfnamefont{J.~A.} \bibnamefont{Caggiano}},
  \bibinfo{author}{\bibfnamefont{B.}~\bibnamefont{Davids}},
  \bibinfo{author}{\bibfnamefont{T.}~\bibnamefont{Glasmacher}},
  \bibnamefont{et~al.}, \bibinfo{journal}{Phys. Rev. Lett.}
  \textbf{\bibinfo{volume}{84}}, \bibinfo{pages}{35} (\bibinfo{year}{2000}).

\bibitem[{\citenamefont{Sauvan et~al.}(2000)\citenamefont{Sauvan, Carstoiu,
  Orr, Ang\'elique, Catford, Clarke, {Mac Cormick}, Curtis, Freer, Gr\'evy
  et~al.}}]{Sau01}
\bibinfo{author}{\bibfnamefont{E.}~\bibnamefont{Sauvan}},
  \bibinfo{author}{\bibfnamefont{F.}~\bibnamefont{Carstoiu}},
  \bibinfo{author}{\bibfnamefont{N.}~\bibnamefont{Orr}},
  \bibinfo{author}{\bibfnamefont{J.}~\bibnamefont{Ang\'elique}},
  \bibinfo{author}{\bibfnamefont{W.}~\bibnamefont{Catford}},
  \bibinfo{author}{\bibfnamefont{N.}~\bibnamefont{Clarke}},
  \bibinfo{author}{\bibfnamefont{M.}~\bibnamefont{{Mac Cormick}}},
  \bibinfo{author}{\bibfnamefont{N.}~\bibnamefont{Curtis}},
  \bibinfo{author}{\bibfnamefont{M.}~\bibnamefont{Freer}},
  \bibinfo{author}{\bibfnamefont{S.}~\bibnamefont{Gr\'evy}},
  \bibnamefont{et~al.}, \bibinfo{journal}{Phys. Lett.}
  \textbf{\bibinfo{volume}{B491}}, \bibinfo{pages}{1} (\bibinfo{year}{2000}).

\bibitem[{\citenamefont{Sauvan et~al.}(2004)\citenamefont{Sauvan, Carstoiu,
  Orr, Winfield, Freer, Ang\'elique, Catford, Clarke, Curtis, Gr\'evy
  et~al.}}]{Setal04}
\bibinfo{author}{\bibfnamefont{E.}~\bibnamefont{Sauvan}},
  \bibinfo{author}{\bibfnamefont{F.}~\bibnamefont{Carstoiu}},
  \bibinfo{author}{\bibfnamefont{N.~A.} \bibnamefont{Orr}},
  \bibinfo{author}{\bibfnamefont{J.~S.} \bibnamefont{Winfield}},
  \bibinfo{author}{\bibfnamefont{M.}~\bibnamefont{Freer}},
  \bibinfo{author}{\bibfnamefont{J.~C.} \bibnamefont{Ang\'elique}},
  \bibinfo{author}{\bibfnamefont{W.~N.} \bibnamefont{Catford}},
  \bibinfo{author}{\bibfnamefont{N.~M.} \bibnamefont{Clarke}},
  \bibinfo{author}{\bibfnamefont{N.}~\bibnamefont{Curtis}},
  \bibinfo{author}{\bibfnamefont{S.}~\bibnamefont{Gr\'evy}},
  \bibnamefont{et~al.}, \bibinfo{journal}{Phys. Rev. C}
  \textbf{\bibinfo{volume}{69}}, \bibinfo{pages}{044603}
  (\bibinfo{year}{2004}).

\bibitem[{\citenamefont{Tostevin et~al.}(2001)\citenamefont{Tostevin, Nunes,
  and Thompson}}]{TNT01}
\bibinfo{author}{\bibfnamefont{J.~A.} \bibnamefont{Tostevin}},
  \bibinfo{author}{\bibfnamefont{F.~M.} \bibnamefont{Nunes}}, \bibnamefont{and}
  \bibinfo{author}{\bibfnamefont{I.~J.} \bibnamefont{Thompson}},
  \bibinfo{journal}{Phys. Rev. C} \textbf{\bibinfo{volume}{63}},
  \bibinfo{pages}{024617} (\bibinfo{year}{2001}).

\bibitem[{\citenamefont{Esbensen et~al.}(2005)\citenamefont{Esbensen, Bertsch,
  and Snover}}]{EBS05}
\bibinfo{author}{\bibfnamefont{H.}~\bibnamefont{Esbensen}},
  \bibinfo{author}{\bibfnamefont{G.~F.} \bibnamefont{Bertsch}},
  \bibnamefont{and} \bibinfo{author}{\bibfnamefont{K.~A.}
  \bibnamefont{Snover}}, \bibinfo{journal}{Phys. Rev. Lett.}
  \textbf{\bibinfo{volume}{94}}, \bibinfo{pages}{042502}
  (\bibinfo{year}{2005}).

\bibitem[{\citenamefont{Capel and Baye}(2005)}]{CB05}
\bibinfo{author}{\bibfnamefont{P.}~\bibnamefont{Capel}} \bibnamefont{and}
  \bibinfo{author}{\bibfnamefont{D.}~\bibnamefont{Baye}},
  \bibinfo{journal}{Phys. Rev. C} \textbf{\bibinfo{volume}{71}},
  \bibinfo{pages}{044609} (\bibinfo{year}{2005}).

\bibitem[{\citenamefont{Summers and Nunes}(2008)}]{SN08}
\bibinfo{author}{\bibfnamefont{N.~C.} \bibnamefont{Summers}} \bibnamefont{and}
  \bibinfo{author}{\bibfnamefont{F.~M.} \bibnamefont{Nunes}},
  \bibinfo{journal}{Phys. Rev. C} \textbf{\bibinfo{volume}{78}},
  \bibinfo{pages}{011601} (\bibinfo{year}{2008}).

\bibitem[{\citenamefont{Flavigny et~al.}(2012)\citenamefont{Flavigny,
  Obertelli, Bonaccorso, Grinyer, Louchart, Nalpas, and Signoracci}}]{Fetal12}
\bibinfo{author}{\bibfnamefont{F.}~\bibnamefont{Flavigny}},
  \bibinfo{author}{\bibfnamefont{A.}~\bibnamefont{Obertelli}},
  \bibinfo{author}{\bibfnamefont{A.}~\bibnamefont{Bonaccorso}},
  \bibinfo{author}{\bibfnamefont{G.~F.} \bibnamefont{Grinyer}},
  \bibinfo{author}{\bibfnamefont{C.}~\bibnamefont{Louchart}},
  \bibinfo{author}{\bibfnamefont{L.}~\bibnamefont{Nalpas}}, \bibnamefont{and}
  \bibinfo{author}{\bibfnamefont{A.}~\bibnamefont{Signoracci}},
  \bibinfo{journal}{Phys. Rev. Lett.} \textbf{\bibinfo{volume}{108}},
  \bibinfo{pages}{252501} (\bibinfo{year}{2012}).

\bibitem[{\citenamefont{Moro et~al.}(arXiv:2004.14612 [nucl-th],
  2020)\citenamefont{Moro, Lay, and G\'omez-Camacho}}]{MLG20}
\bibinfo{author}{\bibfnamefont{A.~M.} \bibnamefont{Moro}},
  \bibinfo{author}{\bibfnamefont{J.~A.} \bibnamefont{Lay}}, \bibnamefont{and}
  \bibinfo{author}{\bibfnamefont{J.}~\bibnamefont{G\'omez-Camacho}}
  (\bibinfo{year}{arXiv:2004.14612 [nucl-th], 2020}).

\bibitem[{\citenamefont{Singh et~al.}(arXiv:2005.05605 [nucl-th],
  2020)\citenamefont{Singh, Matsumoto, and Ogata}}]{SMO20}
\bibinfo{author}{\bibfnamefont{J.}~\bibnamefont{Singh}},
  \bibinfo{author}{\bibfnamefont{T.}~\bibnamefont{Matsumoto}},
  \bibnamefont{and} \bibinfo{author}{\bibfnamefont{K.}~\bibnamefont{Ogata}}
  (\bibinfo{year}{arXiv:2005.05605 [nucl-th], 2020}).

\bibitem[{\citenamefont{Kamimura et~al.}(1986)\citenamefont{Kamimura, Yahiro,
  Iseri, Kameyama, Sakuragi, and Kawai}}]{Kam86}
\bibinfo{author}{\bibfnamefont{M.}~\bibnamefont{Kamimura}},
  \bibinfo{author}{\bibfnamefont{M.}~\bibnamefont{Yahiro}},
  \bibinfo{author}{\bibfnamefont{Y.}~\bibnamefont{Iseri}},
  \bibinfo{author}{\bibfnamefont{H.}~\bibnamefont{Kameyama}},
  \bibinfo{author}{\bibfnamefont{Y.}~\bibnamefont{Sakuragi}}, \bibnamefont{and}
  \bibinfo{author}{\bibfnamefont{M.}~\bibnamefont{Kawai}},
  \bibinfo{journal}{Prog. Theor. Phys. Suppl.} \textbf{\bibinfo{volume}{89}},
  \bibinfo{pages}{1} (\bibinfo{year}{1986}).

\bibitem[{\citenamefont{Yahiro et~al.}(2012)\citenamefont{Yahiro, Ogata,
  Matsumoto, and Minomo}}]{YOMM12}
\bibinfo{author}{\bibfnamefont{M.}~\bibnamefont{Yahiro}},
  \bibinfo{author}{\bibfnamefont{K.}~\bibnamefont{Ogata}},
  \bibinfo{author}{\bibfnamefont{T.}~\bibnamefont{Matsumoto}},
  \bibnamefont{and} \bibinfo{author}{\bibfnamefont{K.}~\bibnamefont{Minomo}},
  \bibinfo{journal}{Progr. Theor. Phys.} \textbf{\bibinfo{volume}{2012}}
  (\bibinfo{year}{2012}).

\bibitem[{\citenamefont{Kido et~al.}(1994)\citenamefont{Kido, Yabana, and
  Suzuki}}]{KYS94}
\bibinfo{author}{\bibfnamefont{T.}~\bibnamefont{Kido}},
  \bibinfo{author}{\bibfnamefont{K.}~\bibnamefont{Yabana}}, \bibnamefont{and}
  \bibinfo{author}{\bibfnamefont{Y.}~\bibnamefont{Suzuki}},
  \bibinfo{journal}{Phys. Rev. C} \textbf{\bibinfo{volume}{50}},
  \bibinfo{pages}{R1276} (\bibinfo{year}{1994}).

\bibitem[{\citenamefont{Esbensen et~al.}(1995)\citenamefont{Esbensen, Bertsch,
  and Bertulani}}]{EBB95}
\bibinfo{author}{\bibfnamefont{H.}~\bibnamefont{Esbensen}},
  \bibinfo{author}{\bibfnamefont{G.}~\bibnamefont{Bertsch}}, \bibnamefont{and}
  \bibinfo{author}{\bibfnamefont{C.}~\bibnamefont{Bertulani}},
  \bibinfo{journal}{Nucl. Phys.} \textbf{\bibinfo{volume}{A581}},
  \bibinfo{pages}{107} (\bibinfo{year}{1995}), ISSN \bibinfo{issn}{0375-9474}.

\bibitem[{\citenamefont{Typel and Wolter}(1999)}]{TW99}
\bibinfo{author}{\bibfnamefont{S.}~\bibnamefont{Typel}} \bibnamefont{and}
  \bibinfo{author}{\bibfnamefont{H.~H.} \bibnamefont{Wolter}},
  \bibinfo{journal}{Z. Naturforsch. Teil A} \textbf{\bibinfo{volume}{54}},
  \bibinfo{pages}{63} (\bibinfo{year}{1999}).

\bibitem[{\citenamefont{Capel et~al.}(2003)\citenamefont{Capel, Baye, and
  Melezhik}}]{CBM03c}
\bibinfo{author}{\bibfnamefont{P.}~\bibnamefont{Capel}},
  \bibinfo{author}{\bibfnamefont{D.}~\bibnamefont{Baye}}, \bibnamefont{and}
  \bibinfo{author}{\bibfnamefont{V.~S.} \bibnamefont{Melezhik}},
  \bibinfo{journal}{Phys. Rev. C} \textbf{\bibinfo{volume}{68}},
  \bibinfo{pages}{014612} (\bibinfo{year}{2003}).

\bibitem[{\citenamefont{Ogata et~al.}(2003)\citenamefont{Ogata, Yahiro, Iseri,
  Matsumoto, and Kamimura}}]{Oetal03}
\bibinfo{author}{\bibfnamefont{K.}~\bibnamefont{Ogata}},
  \bibinfo{author}{\bibfnamefont{M.}~\bibnamefont{Yahiro}},
  \bibinfo{author}{\bibfnamefont{Y.}~\bibnamefont{Iseri}},
  \bibinfo{author}{\bibfnamefont{T.}~\bibnamefont{Matsumoto}},
  \bibnamefont{and} \bibinfo{author}{\bibfnamefont{M.}~\bibnamefont{Kamimura}},
  \bibinfo{journal}{Phys. Rev. C} \textbf{\bibinfo{volume}{68}},
  \bibinfo{pages}{064609} (\bibinfo{year}{2003}).

\bibitem[{\citenamefont{Ogata et~al.}(2006)\citenamefont{Ogata, Hashimoto,
  Iseri, Kamimura, and Yahiro}}]{Oetal06}
\bibinfo{author}{\bibfnamefont{K.}~\bibnamefont{Ogata}},
  \bibinfo{author}{\bibfnamefont{S.}~\bibnamefont{Hashimoto}},
  \bibinfo{author}{\bibfnamefont{Y.}~\bibnamefont{Iseri}},
  \bibinfo{author}{\bibfnamefont{M.}~\bibnamefont{Kamimura}}, \bibnamefont{and}
  \bibinfo{author}{\bibfnamefont{M.}~\bibnamefont{Yahiro}},
  \bibinfo{journal}{Phys. Rev. C} \textbf{\bibinfo{volume}{73}},
  \bibinfo{pages}{024605} (\bibinfo{year}{2006}).

\bibitem[{\citenamefont{Baye et~al.}(2005)\citenamefont{Baye, Capel, and
  Goldstein}}]{BCG05}
\bibinfo{author}{\bibfnamefont{D.}~\bibnamefont{Baye}},
  \bibinfo{author}{\bibfnamefont{P.}~\bibnamefont{Capel}}, \bibnamefont{and}
  \bibinfo{author}{\bibfnamefont{G.}~\bibnamefont{Goldstein}},
  \bibinfo{journal}{Phys. Rev. Lett.} \textbf{\bibinfo{volume}{95}},
  \bibinfo{pages}{082502} (\bibinfo{year}{2005}).

\bibitem[{\citenamefont{Goldstein et~al.}(2006)\citenamefont{Goldstein, Baye,
  and Capel}}]{GBC06}
\bibinfo{author}{\bibfnamefont{G.}~\bibnamefont{Goldstein}},
  \bibinfo{author}{\bibfnamefont{D.}~\bibnamefont{Baye}}, \bibnamefont{and}
  \bibinfo{author}{\bibfnamefont{P.}~\bibnamefont{Capel}},
  \bibinfo{journal}{Phys. Rev. C} \textbf{\bibinfo{volume}{73}},
  \bibinfo{pages}{024602} (\bibinfo{year}{2006}).

\bibitem[{\citenamefont{Glauber}(1959)}]{G59}
\bibinfo{author}{\bibfnamefont{R.~J.} \bibnamefont{Glauber}}, in
  \emph{\bibinfo{booktitle}{Lecture in Theoretical Physics}}, edited by
  \bibinfo{editor}{\bibfnamefont{W.~E.} \bibnamefont{Brittin}}
  \bibnamefont{and} \bibinfo{editor}{\bibfnamefont{L.~G.} \bibnamefont{Dunham}}
  (\bibinfo{publisher}{Interscience}, \bibinfo{address}{New York},
  \bibinfo{year}{1959}), vol.~\bibinfo{volume}{1}, p. \bibinfo{pages}{315}.

\bibitem[{\citenamefont{Hussein and McVoy}(1985)}]{HM85}
\bibinfo{author}{\bibfnamefont{M.~S.} \bibnamefont{Hussein}} \bibnamefont{and}
  \bibinfo{author}{\bibfnamefont{K.~W.} \bibnamefont{McVoy}},
  \bibinfo{journal}{Nucl. Phys. A} \textbf{\bibinfo{volume}{445}},
  \bibinfo{pages}{124} (\bibinfo{year}{1985}).

\bibitem[{\citenamefont{Hencken et~al.}(1996)\citenamefont{Hencken, Bertsch,
  and Esbensen}}]{HBE96}
\bibinfo{author}{\bibfnamefont{K.}~\bibnamefont{Hencken}},
  \bibinfo{author}{\bibfnamefont{G.}~\bibnamefont{Bertsch}}, \bibnamefont{and}
  \bibinfo{author}{\bibfnamefont{H.}~\bibnamefont{Esbensen}},
  \bibinfo{journal}{Phys. Rev. C} \textbf{\bibinfo{volume}{54}},
  \bibinfo{pages}{3043} (\bibinfo{year}{1996}).

\bibitem[{\citenamefont{Hansen and Tostevin}(2003)}]{HT03}
\bibinfo{author}{\bibfnamefont{P.~G.} \bibnamefont{Hansen}} \bibnamefont{and}
  \bibinfo{author}{\bibfnamefont{J.~A.} \bibnamefont{Tostevin}},
  \bibinfo{journal}{Ann. Rev. Nucl. Part. Sc.} \textbf{\bibinfo{volume}{53}},
  \bibinfo{pages}{219} (\bibinfo{year}{2003}).

\bibitem[{\citenamefont{Tostevin et~al.}(2002)\citenamefont{Tostevin, Bazin,
  Brown, Glasmacher, Hansen, Maddalena, Navin, and Sherrill}}]{Tos02}
\bibinfo{author}{\bibfnamefont{J.~A.} \bibnamefont{Tostevin}},
  \bibinfo{author}{\bibfnamefont{D.}~\bibnamefont{Bazin}},
  \bibinfo{author}{\bibfnamefont{B.~A.} \bibnamefont{Brown}},
  \bibinfo{author}{\bibfnamefont{T.}~\bibnamefont{Glasmacher}},
  \bibinfo{author}{\bibfnamefont{P.~G.} \bibnamefont{Hansen}},
  \bibinfo{author}{\bibfnamefont{V.}~\bibnamefont{Maddalena}},
  \bibinfo{author}{\bibfnamefont{A.}~\bibnamefont{Navin}}, \bibnamefont{and}
  \bibinfo{author}{\bibfnamefont{B.~M.} \bibnamefont{Sherrill}},
  \bibinfo{journal}{Phys. Rev. C} \textbf{\bibinfo{volume}{66}},
  \bibinfo{pages}{024607} (\bibinfo{year}{2002}).

\bibitem[{\citenamefont{Yahiro et~al.}(2011)\citenamefont{Yahiro, Ogata, and
  Minomo}}]{YOM11}
\bibinfo{author}{\bibfnamefont{M.}~\bibnamefont{Yahiro}},
  \bibinfo{author}{\bibfnamefont{K.}~\bibnamefont{Ogata}}, \bibnamefont{and}
  \bibinfo{author}{\bibfnamefont{K.}~\bibnamefont{Minomo}},
  \bibinfo{journal}{Progr. Theor. Phys.} \textbf{\bibinfo{volume}{126}},
  \bibinfo{pages}{167} (\bibinfo{year}{2011}).

\bibitem[{\citenamefont{Minomo et~al.}(2014)\citenamefont{Minomo, Matsumoto,
  Egashira, Ogata, and Yahiro}}]{Metal14}
\bibinfo{author}{\bibfnamefont{K.}~\bibnamefont{Minomo}},
  \bibinfo{author}{\bibfnamefont{T.}~\bibnamefont{Matsumoto}},
  \bibinfo{author}{\bibfnamefont{K.}~\bibnamefont{Egashira}},
  \bibinfo{author}{\bibfnamefont{K.}~\bibnamefont{Ogata}}, \bibnamefont{and}
  \bibinfo{author}{\bibfnamefont{M.}~\bibnamefont{Yahiro}},
  \bibinfo{journal}{Phys. Rev. C} \textbf{\bibinfo{volume}{90}},
  \bibinfo{pages}{027601} (\bibinfo{year}{2014}).

\bibitem[{\citenamefont{Fukui et~al.}(2014)\citenamefont{Fukui, Ogata, and
  Capel}}]{FOC14}
\bibinfo{author}{\bibfnamefont{T.}~\bibnamefont{Fukui}},
  \bibinfo{author}{\bibfnamefont{K.}~\bibnamefont{Ogata}}, \bibnamefont{and}
  \bibinfo{author}{\bibfnamefont{P.}~\bibnamefont{Capel}},
  \bibinfo{journal}{Phys.~Rev.~C} \textbf{\bibinfo{volume}{90}},
  \bibinfo{pages}{034617} (\bibinfo{year}{2014}).

\bibitem[{\citenamefont{Baye and Capel}(2012)}]{BC12}
\bibinfo{author}{\bibfnamefont{D.}~\bibnamefont{Baye}} \bibnamefont{and}
  \bibinfo{author}{\bibfnamefont{P.}~\bibnamefont{Capel}}, in
  \emph{\bibinfo{booktitle}{Clusters in Nuclei, Vol. 2}}, edited by
  \bibinfo{editor}{\bibfnamefont{C.}~\bibnamefont{Beck,}}
\bibnamefont{Lecture Notes in Physics Vol. 848}
  (\bibinfo{publisher}{Springer}, \bibinfo{address}{Heidelberg},
  \bibinfo{year}{2012}), p. \bibinfo{volume}{121}.

\bibitem[{\citenamefont{Hashimoto et~al.}(2011)\citenamefont{Hashimoto, Yahiro,
  Ogata, Minomo, and Chiba}}]{Hetal11}
\bibinfo{author}{\bibfnamefont{S.}~\bibnamefont{Hashimoto}},
  \bibinfo{author}{\bibfnamefont{M.}~\bibnamefont{Yahiro}},
  \bibinfo{author}{\bibfnamefont{K.}~\bibnamefont{Ogata}},
  \bibinfo{author}{\bibfnamefont{K.}~\bibnamefont{Minomo}}, \bibnamefont{and}
  \bibinfo{author}{\bibfnamefont{S.}~\bibnamefont{Chiba}},
  \bibinfo{journal}{Phys. Rev. C} \textbf{\bibinfo{volume}{83}},
  \bibinfo{pages}{054617} (\bibinfo{year}{2011}).

\bibitem[{\citenamefont{Hebborn and Baye}(2020)}]{HB20}
\bibinfo{author}{\bibfnamefont{C.}~\bibnamefont{Hebborn}} \bibnamefont{and}
  \bibinfo{author}{\bibfnamefont{D.}~\bibnamefont{Baye}},
  \bibinfo{journal}{Phys. Rev. C} \textbf{\bibinfo{volume}{101}},
  \bibinfo{pages}{054609} (\bibinfo{year}{2020}).

\bibitem[{\citenamefont{Margueron et~al.}(2003)\citenamefont{Margueron,
  Bonaccorso, and Brink}}]{MBB03}
\bibinfo{author}{\bibfnamefont{J.}~\bibnamefont{Margueron}},
  \bibinfo{author}{\bibfnamefont{A.}~\bibnamefont{Bonaccorso}},
  \bibnamefont{and} \bibinfo{author}{\bibfnamefont{D.~M.} \bibnamefont{Brink}},
  \bibinfo{journal}{Nucl. Phys. A} \textbf{\bibinfo{volume}{720}},
  \bibinfo{pages}{337} (\bibinfo{year}{2003}).

\bibitem[{\citenamefont{Capel et~al.}(2008)\citenamefont{Capel, Baye, and
  Suzuki}}]{CBS08}
\bibinfo{author}{\bibfnamefont{P.}~\bibnamefont{Capel}},
  \bibinfo{author}{\bibfnamefont{D.}~\bibnamefont{Baye}}, \bibnamefont{and}
  \bibinfo{author}{\bibfnamefont{Y.}~\bibnamefont{Suzuki}},
  \bibinfo{journal}{Phys. Rev. C} \textbf{\bibinfo{volume}{78}},
  \bibinfo{pages}{054602} (\bibinfo{year}{2008}).

\bibitem[{\citenamefont{Capel et~al.}(2018)\citenamefont{Capel, Phillips, and
  Hammer}}]{CPH18}
\bibinfo{author}{\bibfnamefont{P.}~\bibnamefont{Capel}},
  \bibinfo{author}{\bibfnamefont{D.~R.} \bibnamefont{Phillips}},
  \bibnamefont{and} \bibinfo{author}{\bibfnamefont{H.-W.}
  \bibnamefont{Hammer}}, \bibinfo{journal}{Phys. Rev. C}
  \textbf{\bibinfo{volume}{98}}, \bibinfo{pages}{034610}
  (\bibinfo{year}{2018}).

\bibitem[{\citenamefont{Hebborn and Capel}(2019)}]{HC19}
\bibinfo{author}{\bibfnamefont{C.}~\bibnamefont{Hebborn}} \bibnamefont{and}
  \bibinfo{author}{\bibfnamefont{P.}~\bibnamefont{Capel}},
  \bibinfo{journal}{Phys. Rev. C} \textbf{\bibinfo{volume}{100}},
  \bibinfo{pages}{054607} (\bibinfo{year}{2019}).

\bibitem[{\citenamefont{Bertulani et~al.}(2002)\citenamefont{Bertulani, Hammer,
  and van Kolck}}]{BHvk02}
\bibinfo{author}{\bibfnamefont{C.~A.} \bibnamefont{Bertulani}},
  \bibinfo{author}{\bibfnamefont{H.-W.} \bibnamefont{Hammer}},
  \bibnamefont{and} \bibinfo{author}{\bibfnamefont{U.}~\bibnamefont{van
  Kolck}}, \bibinfo{journal}{Nucl. Phys. A} \textbf{\bibinfo{volume}{712}},
  \bibinfo{pages}{37} (\bibinfo{year}{2002}).

\bibitem[{\citenamefont{Bedaque et~al.}(2003)\citenamefont{Bedaque, Hammer, and
  van Kolck}}]{BHvk03}
\bibinfo{author}{\bibfnamefont{P.~F.} \bibnamefont{Bedaque}},
  \bibinfo{author}{\bibfnamefont{H.-W.} \bibnamefont{Hammer}},
  \bibnamefont{and} \bibinfo{author}{\bibfnamefont{U.}~\bibnamefont{van
  Kolck}}, \bibinfo{journal}{Phys. Lett. B} \textbf{\bibinfo{volume}{569}},
  \bibinfo{pages}{159} (\bibinfo{year}{2003}).

\bibitem[{\citenamefont{Hammer et~al.}(2017)\citenamefont{Hammer, Ji, and
  Phillips}}]{HJP17}
\bibinfo{author}{\bibfnamefont{H.-W.} \bibnamefont{Hammer}},
  \bibinfo{author}{\bibfnamefont{C.}~\bibnamefont{Ji}}, \bibnamefont{and}
  \bibinfo{author}{\bibfnamefont{D.~R.} \bibnamefont{Phillips}},
  \bibinfo{journal}{J. Phys. G} \textbf{\bibinfo{volume}{44}},
  \bibinfo{pages}{103002} (\bibinfo{year}{2017}).

\bibitem[{\citenamefont{Calci et~al.}(2016)\citenamefont{Calci, Navr\'atil,
  Roth, Dohet-Eraly, Quaglioni, and Hupin}}]{Cetal16}
\bibinfo{author}{\bibfnamefont{A.}~\bibnamefont{Calci}},
  \bibinfo{author}{\bibfnamefont{P.}~\bibnamefont{Navr\'atil}},
  \bibinfo{author}{\bibfnamefont{R.}~\bibnamefont{Roth}},
  \bibinfo{author}{\bibfnamefont{J.}~\bibnamefont{Dohet-Eraly}},
  \bibinfo{author}{\bibfnamefont{S.}~\bibnamefont{Quaglioni}},
  \bibnamefont{and} \bibinfo{author}{\bibfnamefont{G.}~\bibnamefont{Hupin}},
  \bibinfo{journal}{Phys. Rev. Lett.} \textbf{\bibinfo{volume}{117}},
  \bibinfo{pages}{242501} (\bibinfo{year}{2016}).

\bibitem[{\citenamefont{Moschini and Capel}(2019)}]{MC19}
\bibinfo{author}{\bibfnamefont{L.}~\bibnamefont{Moschini}} \bibnamefont{and}
  \bibinfo{author}{\bibfnamefont{P.}~\bibnamefont{Capel}},
  \bibinfo{journal}{Phys. Lett.} \textbf{\bibinfo{volume}{B790}},
  \bibinfo{pages}{367} (\bibinfo{year}{2019}).

\bibitem[{\citenamefont{Yang and Capel}(2018)}]{YC18}
\bibinfo{author}{\bibfnamefont{J.}~\bibnamefont{Yang}} \bibnamefont{and}
  \bibinfo{author}{\bibfnamefont{P.}~\bibnamefont{Capel}},
  \bibinfo{journal}{Phys. Rev. C} \textbf{\bibinfo{volume}{98}},
  \bibinfo{pages}{054602} (\bibinfo{year}{2018}).

\bibitem[{\citenamefont{Hebborn}(2020)}]{PhDHebborn}
\bibinfo{author}{\bibfnamefont{C.}~\bibnamefont{Hebborn}}, Ph.D. thesis,
  \bibinfo{school}{Universit\'e libre de Bruxelles},
  \bibinfo{address}{Brussels} (\bibinfo{year}{2020}),
  \urlprefix\url{http://hdl.handle.net/2013/ULB-DIPOT:oai:dipot.ulb.ac.be:2013/312495}.

\bibitem[{\citenamefont{Hebborn and Capel}(2021)}]{HC21}
\bibinfo{author}{\bibfnamefont{C.}~\bibnamefont{Hebborn}} \bibnamefont{and}
  \bibinfo{author}{\bibfnamefont{P.}~\bibnamefont{Capel}},
  \bibinfo{howpublished}{in preparation} (\bibinfo{year}{2021}).

\end{thebibliography}

\end{document}